\documentclass[10pt,twocolumn,a4paper] {IEEEtran} 
\usepackage[margin=15mm,bottom=20mm]{geometry}
\usepackage[utf8]{inputenc}
\usepackage{amsfonts}
\usepackage{amsmath}
\usepackage{multicol}
\usepackage{amssymb}
\usepackage{graphicx}
\usepackage{mathtools}
\usepackage{xcolor}

\usepackage{cite}
\usepackage{breqn}

\usepackage{units}

\begin{document}

\title{
Understanding the Basis of Graph Signal Processing via an Intuitive Example-Driven Approach}

\author{Ljubi\v{s}a Stankovi\'{c}, 
Danilo Mandic,
 Milo\v{s} Dakovi\'{c},
Ilya Kisil,
Ervin Sejdi\'{c},
Anthony G. Constantinides}


\maketitle

\setcounter{tocdepth}{3}


\section{Scope}

Graphs are irregular structures which naturally account for data integrity, however, traditional approaches have been established outside Signal Processing, and largely focus on analyzing the underlying graphs rather than signals on graphs.
 Given the rapidly increasing availability of multisensor and multinode measurements, likely recorded on irregular or ad-hoc grids, it would be extremely advantageous to analyze such structured data as graph signals and thus benefit from the ability of graphs to incorporate spatial awareness of the sensing locations, physical intuition and sensor importance, and the ease of local versus global sensor association.  The aim of this Lecture Note is therefore to establish a common language between graph signals, defined on irregular signal domains, and some of the most fundamental paradigms in DSP, such as spectral analysis of multichannel signals, system transfer function, digital filter design, parameter estimation, and optimal denoising.  

This is achieved through a physically meaningful and intuitive real-world example of geographically distributed multisensor temperature estimation.
A similar spatial multisensor arrangement is already widely used in Signal Processing curricula to introduce minimum variance estimators and Kalman filters, and by adopting this framework we facilitate a seamless integration of graph theory into the curriculum of existing DSP courses. By bridging the gap between standard approaches and graph signal processing, we also show that standard methods can be thought of as special cases of their graph counterparts, evaluated on line graphs. It is hoped that our approach would not only help to demystify graph theoretic approaches in education but it would also empower practitioners  and researchers to explore a whole host of otherwise prohibitive modern applications.

\section{Relevance}

 In classical Signal Processing, the signal domain is determined by equidistant time instants or by a set of spatial sensing points on a uniform grid. However, increasingly the actual data sensing domain may not even be related to the physical dimensions of time and/or  space, and it typically does exhibit various forms of regularity. For example, in social or web-related networks, the sensing points and their connectivity pertain to specific objects/nodes and topology of their links. It should be noted that  even for the data acquired in well defined time and space domains,  the introduction of new relations between the signal samples, through graphs, may yield new insights into the analysis and provide enhanced data processing (e.g., based on local similarity neighborhoods). The advantage of graphs over classical data domains is that graphs account naturally for irregular data relations in the problem definition, together with the corresponding data connectivity in the analysis.

Indeed, Graph Signal and Information Processing is particularly well suited to making sense from data acquired over irregular data domains, which can be achieved, for example, by leveraging intuitions developed on Euclidean domains,  by employing analogies with other irregular domains such as polygon meshes and manifolds, or learning the mutual connectivity from available sets of data. In many emerging applications, e.g., Big Data, this also introduces a number of new challenges:
\begin{itemize}
	\item 
	Basic concepts must be revisited in order to accommodate structured but often incomplete information, 
	\item New physically meaningful frameworks, specifically tailored for heterogeneous data sources, are required, and
	\item Trade-offs between performance and numerical requirements are a prerequisite when operating in real-time.
\end{itemize}

The common language and enhanced intuition between the  graph  approaches and their standard counterparts, illuminated in this article through  the relationships between the vertex and time domains,  may be naturally generalized to address the above challenges and spur further developments in the curricula on Statistical Signal Processing, Graph Signal Processing, and Big Data.

\section{Prerequisites}
This Lecture Note assumes a basic knowledge of Linear Algebra and Digital Signal Processing.  
 
\section{History of Graph Theoretic Application}
Graph theory, as a branch of mathematics, has existed for almost three centuries. 
The beginning of graph theory applications in electrical engineering dates back to the mid-XIX century and the definition of Kirchoff's laws. 
Owing to their inherent ``spatial awareness", graph models have since become a de facto standard for data analysis across the science and engineering areas, including chemistry,  operational research, social networks, and computer sciences. 

A systematic account of graph theory as an optimization tool can be attributed to the seminal book by Nicos Christofides of Imperial College London, published in 1975 \cite{NC}. Soon after gaining prominence in general optimization, it was very natural to explore the application of graph theory in signal processing and related areas \cite{TC}. Indeed, perhaps the first lecture course to teach graph theory to then emerging communication networks and channel coding student cohort was introduced by the author Anthony Constantinides in 1970s. This helped to establish and formalize the connections between general optimization and the topology of a communication network, and has spurred further applications in image processing \cite{AC}.

After a relative lull over the  next two decades, current developments in graph theory owe their prominence to the emergence of modern data sources, such as large-scale sensor and social networks, which inherently provide rich underlying physical, social, and geographic structures that require new ways to {\color{black} establish statistical inference, leading to data processing on graphs, within a new fast maturing field of Graph Signal Processing, \cite{JK,Maura1,Maura2,Chu,shuman2013may,DM,VF}. }

\section{Problem Statement: An Illustrative Example}
 	{\color{black}Graphs and Graph Signal Processing represent quite a general mathematical formalism which, albeit different from classic concepts, does admit the development of graph-domain counterparts of well established DSP paradigms. It would therefore be valuable to introduce such a general concept in an inductive and intuitive way, through a  simple, general enough and well-understood example regarding a commonly considered topic in classical DSP.} 
 	
\begin{figure}[t]
\centering
\includegraphics[  width=8.7cm]{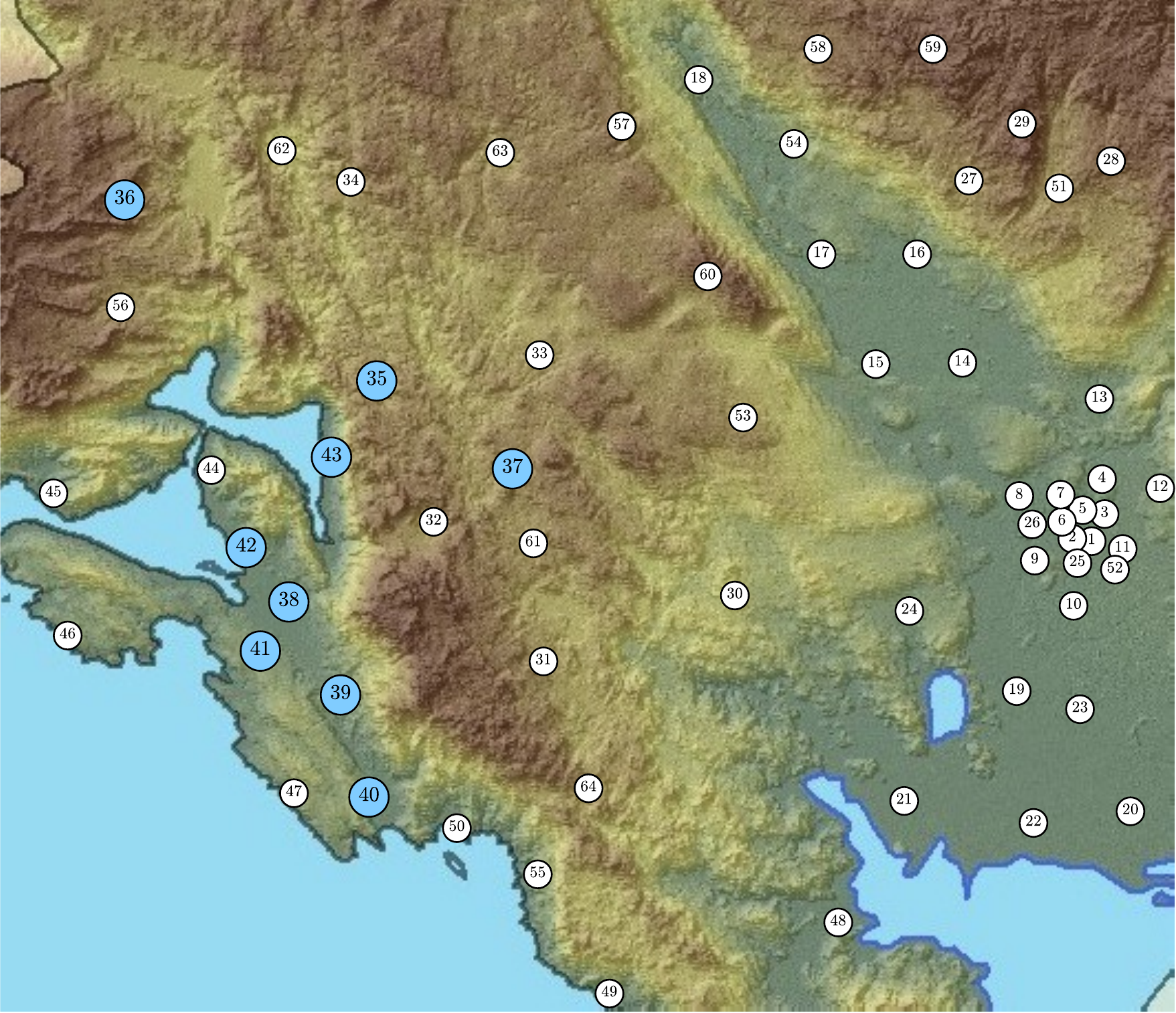}\hspace{5mm} a)

\vspace{5mm}

\includegraphics[  width=8.7cm]{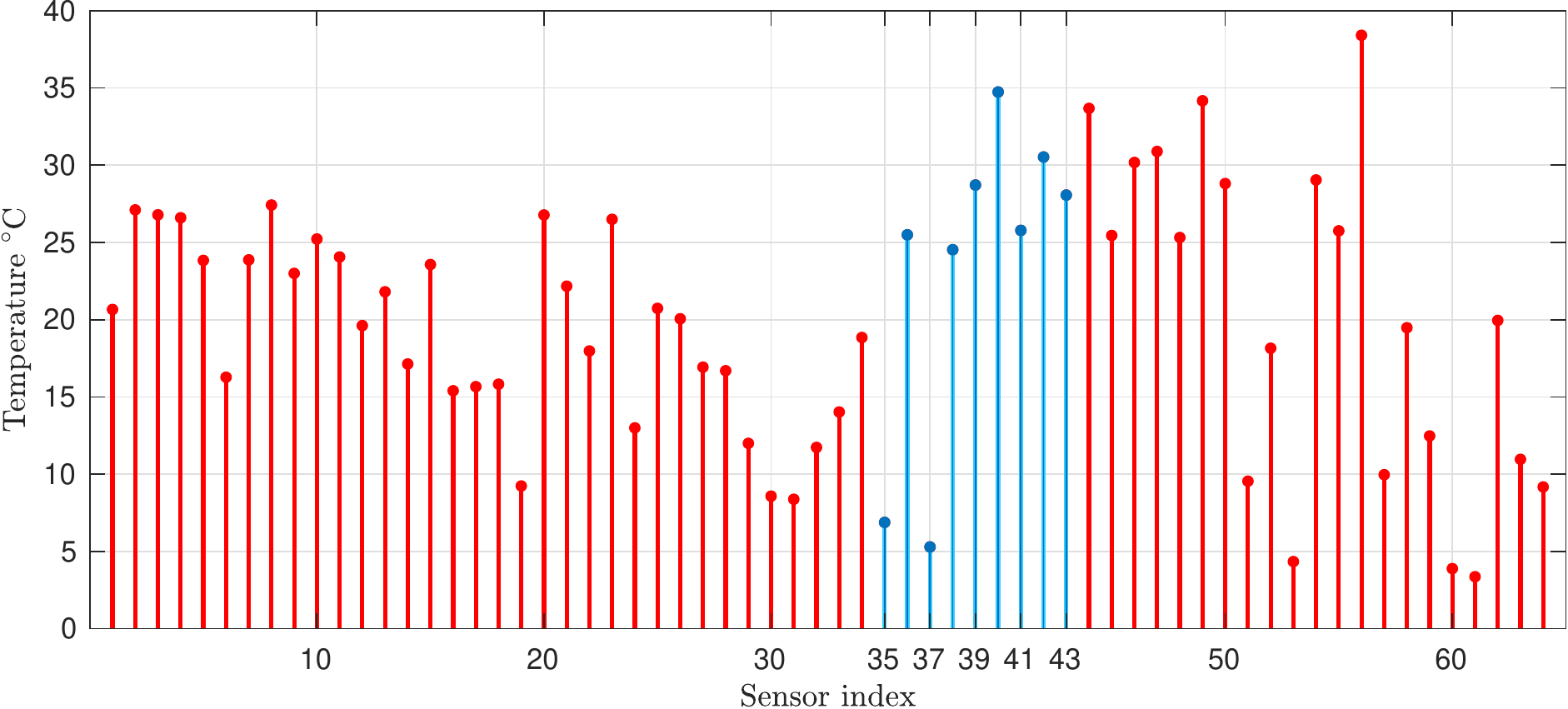}

\vspace{2mm}
\hspace*{4mm}
\includegraphics[width=8.3cm]{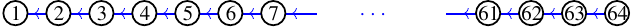}\hspace{5mm} b)

\caption{Temperature sensing as a classical signal processing problem.
a) Sensing locations in a geographic region along the Adriatic area.  b) Temperatures measured at $N=64$ sensing locations (top). In standard signal processing, the spatial sensor index is used for the horizontal axis and serves as the signal domain. This domain can be interpreted as a directed line graph (bottom). Observe the lack of physical intuition, as for example, sensor $37$ (mountains) is followed by sensor $38$ (coast), with drastic difference in temperature.
}
\label{fig:Prva_slika_a}
\end{figure}

 	To this end consider a multi-sensor setup, shown in Fig. \ref{fig:Prva_slika_a}, for measuring a temperature field in a known geographical region; such a set-up is typically used in the context of minimum-variance estimators and Kalman filters. The temperature sensing locations are chosen according to the significance of a particular geographic area to local users, with $N=64$ sensing points in total, as shown in Fig. \ref{fig:Prva_slika_a}a). The temperature field is denoted by $\{x(n)\}$ and a snapshot of its values is given in Fig. \ref{fig:Prva_slika_a}b). 
 Each measured sensor signal can then be mathematically expressed as \begin{equation}
 x(n)=s(n)+\varepsilon(n),\label{nsig}
 \end{equation}
  where $s(n)$ is the true temperature  that would have be obtained in  ideal measuring conditions and $\varepsilon(n)$ comprises the adverse effects of the local environment on sensor readings or faulty sensor activity, and is referred to as ``noise" in the sequel. For illustrative purposes, in our study $\varepsilon(n)$ was modeled as a realization of white, zero-mean, Gaussian process, with standard deviation $\sigma_\varepsilon=4$. It was added to the signal, $s(n)$, to yield the signal-to-noise ratio in $\{x(n)\}$ of $SNR_0=14.2$ dB.  

\textit{Remark 1:} Classical signal processing requires an arrangement of the quintessentially spatial temperature samples in Fig. \ref{fig:Prva_slika_a}a) into a {\color{black}line} structure shown in Fig. \ref{fig:Prva_slika_a}b). Obviously, such ``lexicographic" ordering is not amenable to exploiting the spatial information related to the actual sensor arrangement, dictated by the terrain. For example, this renders classical analyses of this temperature field inapplicable (or at best suboptimal), as the performance critically depends on the chosen sensor ordering scheme. This exemplifies that even a most routine temperature measurement  setup requires a more complex estimation structure than the simple {\color{black} line} one corresponding to the classical signal processing framework, shown in Fig. \ref{fig:Prva_slika_a}b).  

To introduce a ``situation-aware" noise reduction scheme for the temperature field in Fig. \ref{fig:Prva_slika_a}, we proceed to explore  a graph-theoretic framework to this problem, starting from a local signal average operator. In classical Signal Processing this can be achieved through a moving average operator, e.g., through averaging across the neighboring data samples, or equivalently neighboring nodes, as in the {\color{black} line} graph in Fig. \ref{fig:Prva_slika_a}b), and  for each sensing point. Physically, such local neighborhood should indeed include close neighboring sensing points but which also exhibit similar meteorological properties defined by the distance, altitude difference, and other terrain properties. In other words, since the sensor network in Fig. \ref{fig:Prva_slika_a} measures a set of related temperatures from irregularly spaced sensors, \textbf{an effective estimation strategy should include domain knowledge -- not possible to achieve with standard DSP} ({\color{black}line graph}).   

\begin{figure}[t]
	\centering
	\includegraphics[  width=8.5cm]{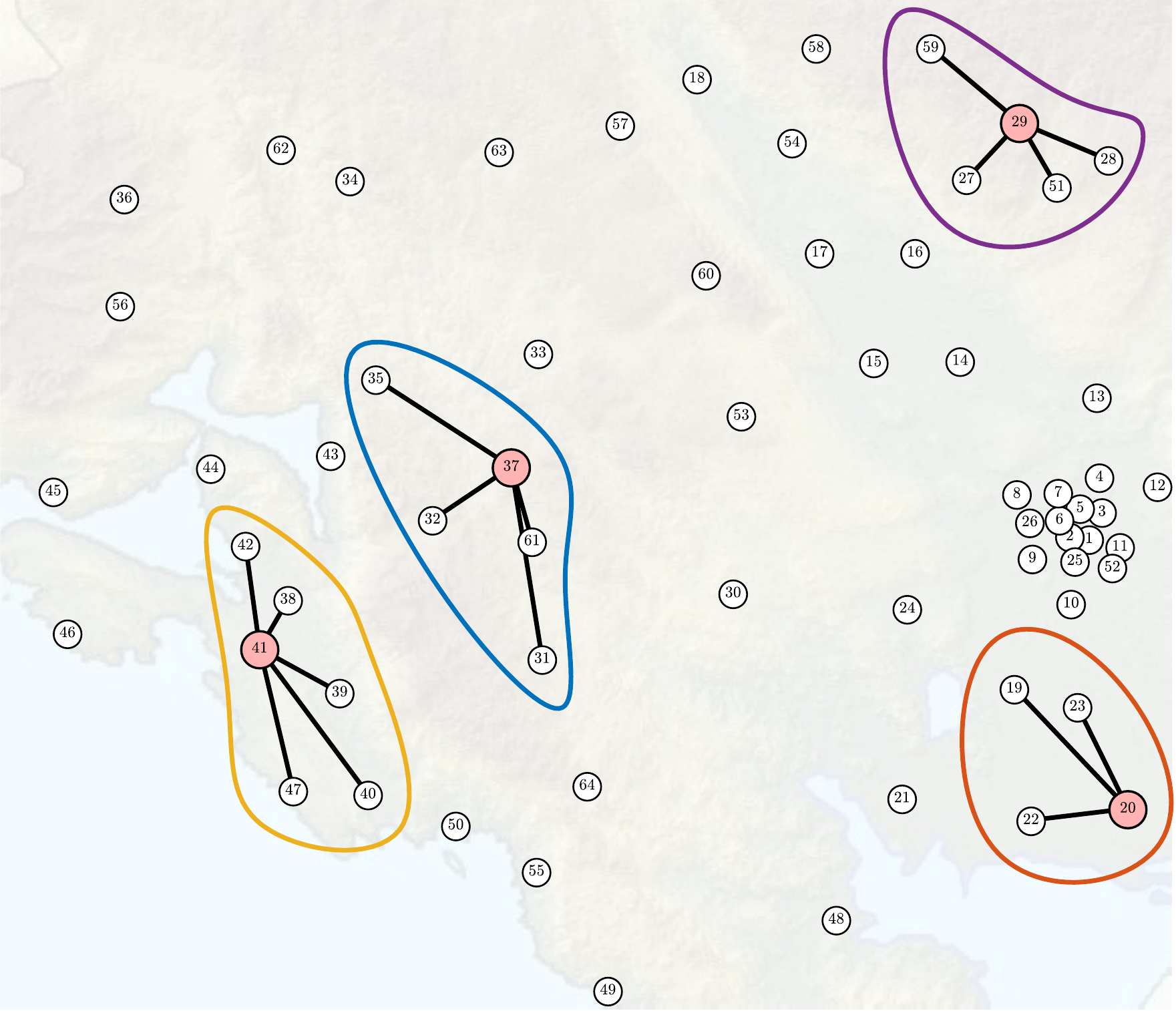}a)
	\hfill
	
	\vspace{5mm}
		
	\includegraphics[  width=8.5cm]{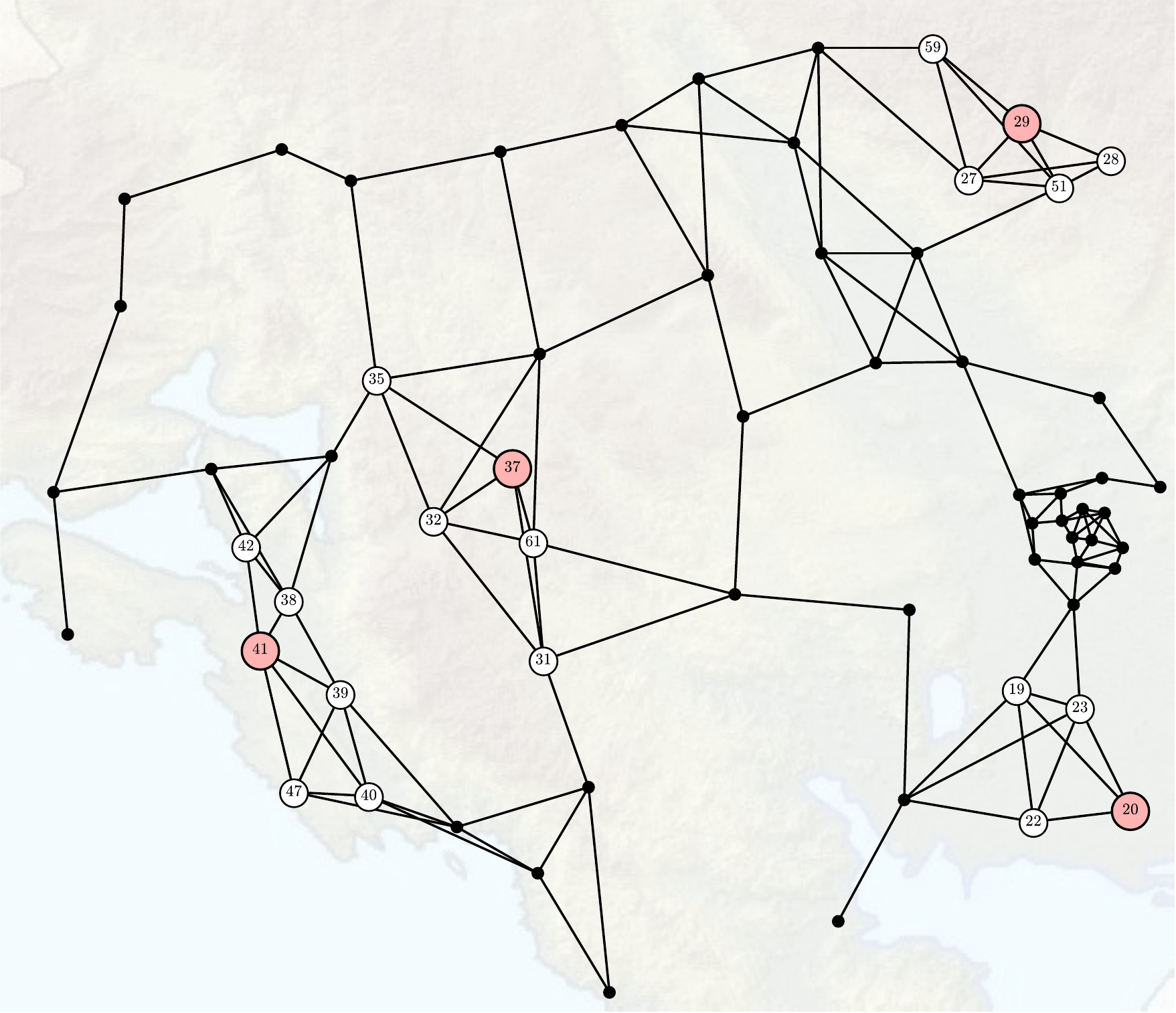}b)
	
	\caption{Temperature setup as a domain-aware graph signal processing problem. a) Local neighborhood for the sensing points  $n=20$, $29$, $37$, and $41$. These neighborhoods are chosen using ``domain knowledge", dictated by local terrain and by taking into account the distance and altitude of sensors.  Neighboring sensors for each of these sensing locations (vertices) are chosen in a physically meaningful way and their relation is indicated by the connectivity lines, called edges. b) Local neighborhoods for all sensing vertices, presented in a graph form.}
	\label{fig:Prva_slika_b}
\end{figure}

Consider the local neighborhoods for the sensing points $n=20$, $29$, $37$, and $41$, shown in Fig. \ref{fig:Prva_slika_b}a). The cumulative temperature for each sensing point is then given by
$$y(n)=\sum_{m \text{ at  and around } n  } x(m),$$ 
so that the local average temperature for a sensing point $n$ may be easily obtained by dividing the cumulative temperature, $y(n)$, with the number of included sensing points.
For example, for the sensing points $n=20$ and $n=37$, presented in Fig. \ref{fig:Prva_slika_b}a), the ``domain knowledge aware" local estimation takes the form
\begin{align}
y(20) & =x(20)+x(19)+x(22)+x(23) \label{20VA}\\
y(37) & = x(37)+x(32)+x(33)+x(35)+x(61). \label{37VA}
\end{align}
For convenience, the full set of relations among the sensing points can now be arranged into the matrix form, to give
\begin{equation}
\mathbf{y}=\mathbf{x}+\mathbf{A}\mathbf{x}, \label{1}
\end{equation} 
where the matrix $\mathbf{A}$ indicates the connectivity structure of the neighboring sensing locations that should be involved in the calculation for each $y(n)$. 
The matrix  $\mathbf{A}$ is therefore referred to as the \textbf{connectivity or adjacency matrix} of a graph. Its elements are either $1$ (if the corresponding vertices are related) or $0$ (if they are not related). Fig. \ref{fig:Prva_slika_b}b) shows the sensing locations with the corresponding connectivity for the temperature estimation scenario in Fig. \ref{fig:Prva_slika_b}a). From (\ref{20VA}) we can observe, for example, that the $20$th row of the adjacency matrix $\mathbf{A}$  will have all zero elements, except for $A_{20,19}=1$, $A_{20,22}=1$, and $A_{20,23}=1$ (for more detail see the electronic supplement). 

\begin{figure}[t]
\centering
\includegraphics[  width=7.5cm]{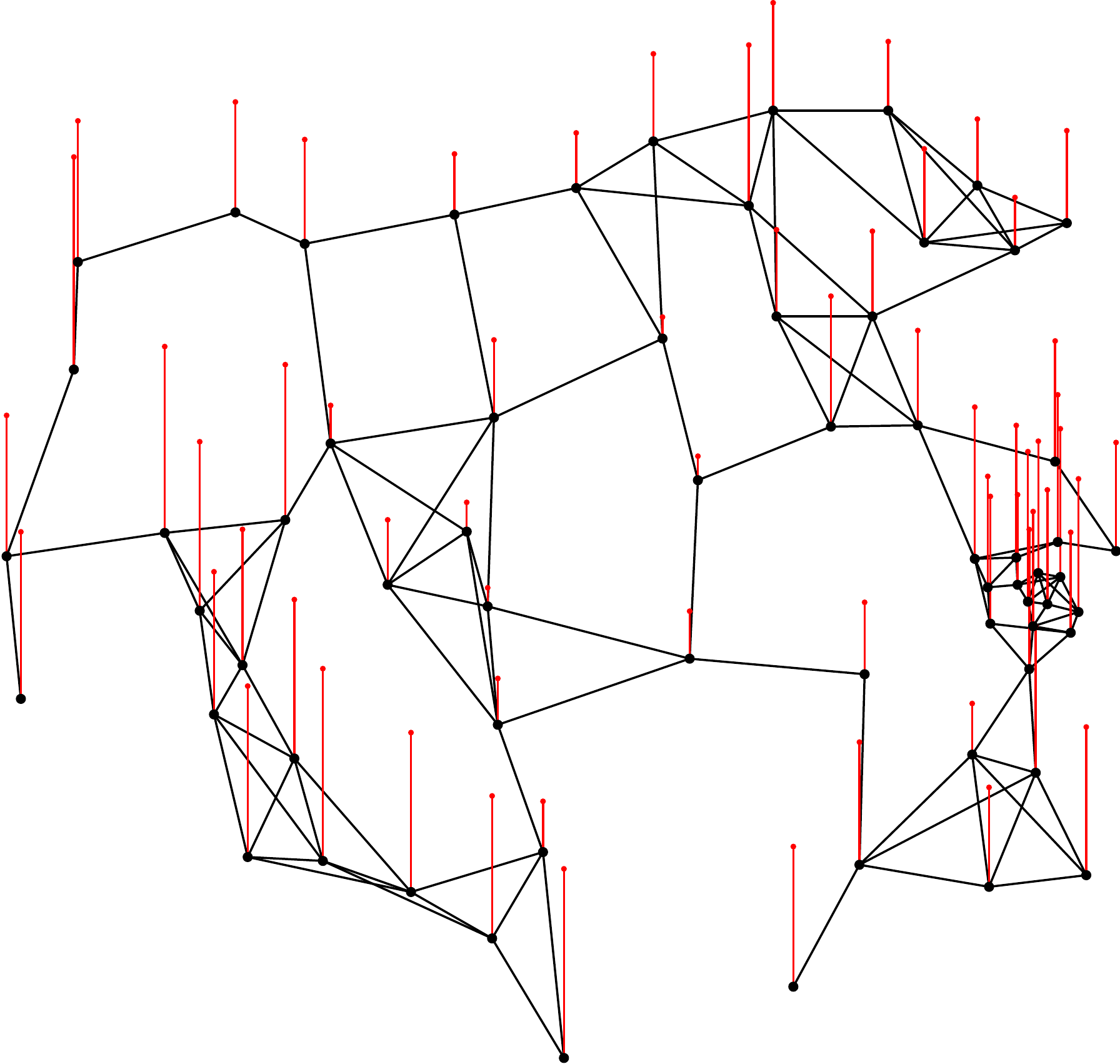} a)  

\hfill

\vspace{5mm}

\includegraphics[  width=8.2cm]{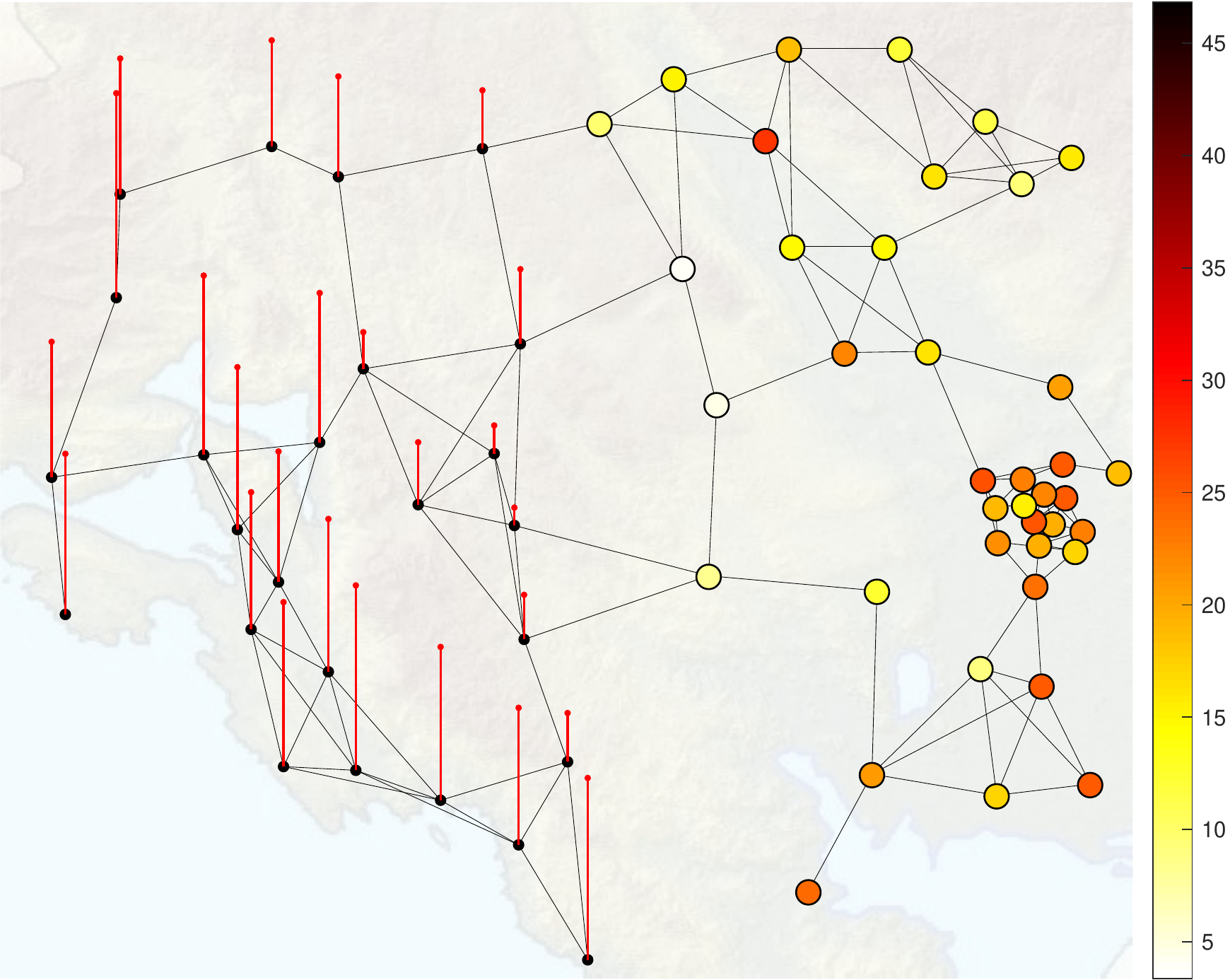} b)
\caption{From a multi-sensor measurement to a graph signal. a) The temperature field is represented on a graph that combines spatially unaware measurements in Fig. \ref{fig:Prva_slika_a}b) and the physically relevant graph topology in Fig. \ref{fig:Prva_slika_b}b).  b) The graph signal intensity may also be designated by the vertex color, as in the right half of the panel.}
\label{fig:Prva_slika_e}.
\end{figure}

This simple real-world example can be interpreted within the graph signal processing framework as follows: 
\begin{itemize}
	\item The sensing points where the signal is measured are designated as the \textbf{graph vertices}, see Fig. \ref{fig:Prva_slika_a},
	\item The vertex-to-vertex lines indicating the connectivity among the sensing points are called the \textbf{graph edges}, 
	\item The vertices and edges form \textbf{a graph}, as in Fig. \ref{fig:Prva_slika_b}b), a new and very structurally rich signal domain, 
	\item  The graph, rather than a standard vector of sensing points, is then used for analyzing and processing data, as it is equipped with spatial and physical awareness,  
	\item The measured temperatures are now interpreted as \textbf{signal samples on graph}, as shown in Fig. \ref{fig:Prva_slika_e}, 	
	\item   Similar to traditional signal processing, this new \textbf{graph signal} may have many realizations on the same graph and may include noise,
	\item Through relation (\ref{1}), we have therefore introduced a simple \textbf{graph system} for physically and spatially aware signal averaging (a linear first-order graph system).
\end{itemize}

To emphasize our trust in a particular sensor and to model mutual sensor  relevance,  a weighting scheme may be imposed on the edges (connectivity) between the sensing points, in the form
\begin{equation}
y(n)=x(n)+\sum_{m\ne n} W_{nm}x(m). \label{2}
\end{equation}
The weight $W_{nm}$ indicates the strength of the coupling between signal values at the sensing points $n$ and $m$; its value is zero if the points $n$ and $m$ are not related and for $n=m$. We have now arrived at a weighted graph, whereby each edge has an associated weight, $ W_{nm}$, which adds a ``mutual sensor relevance" information to the already established ``spatial awareness" modeled by the edges.  In our example, a matrix form of a weighted cumulative graph signal now becomes 
\begin{equation}
\mathbf{y}=\mathbf{x}+\mathbf{W}\mathbf{x}. \label{3}
\end{equation}
This equips graph signal models with additional flexibility. In order to produce unbiased estimates, instead of the cumulative sums in (\ref{1}) and (\ref{2}), the weighting coefficients within the estimate for each $y(n)$ should sum up to unity. This may be achieved through a  normalized form of (\ref{3}), given by
\begin{equation}
\mathbf{y}=\frac{1}{2}(\mathbf{x}+{\mathbf{D}^{-1}}\mathbf{W}\mathbf{x}),\label{AWjed}
\end{equation}
where the elements of the diagonal normalization matrix, $\mathbf{D}$, called \textbf{the degree matrix}, are $D_{nn}=\sum_m W_{nm}$ {\color{black} and ${\mathbf{D}^{-1}}\mathbf{W}$ is referred to as a \textit{random walk} weight matrix.}  
When this simple normalized first-order system is employed to filter the original noisy signal from Fig. \ref{fig:Prva_slika_e}, an improvement of  $6$ dB over the original signal-to-noise ratio, $SNR_0=14.2$ dB, is achieved.

Another important operator for graph signal processing is the \textbf{graph Laplacian}, $\mathbf{L}$, which is defined as
$$ \mathbf{L}=\mathbf{D}-\mathbf{W}. $$

\begin{figure*}[tbp]
\centering
\fboxsep=1pt
\fbox{\includegraphics[]{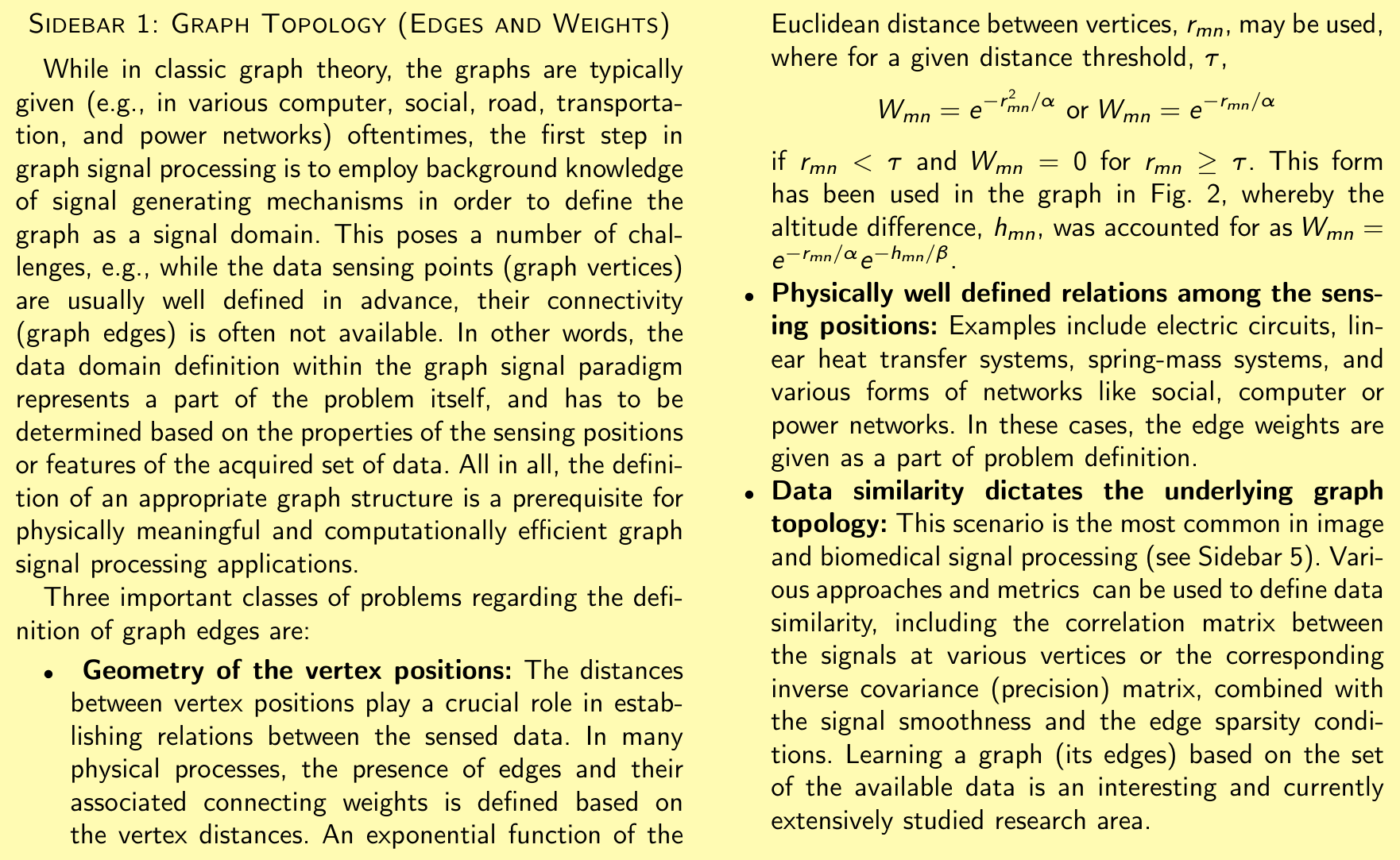}}
\end{figure*}

\bigskip
\textit{Remark 2:} A graph is fully specified  by the set of its vertices and their connectivity scheme (designated by edges). The edges may be defined by the adjacency matrix, $\mathbf{A}$, with $A_{mn} \in \{0,1\}$, for unweighted graphs or by the ``connectivity strength" weighting matrix, $\mathbf{W}$, with $W_{mn} \in \mathbb{R}^+$, for weighted graphs. The degree matrix, $\mathbf{D}$, and the Laplacian matrix,  $\mathbf{L}$, with $L_{mn} \in \mathbb{R}$,  are defined using the adjacency/weighting matrix. When the relations between all pairs of vertices are mutually symmetric, then all the matrices involved are also symmetric, and such graphs are called \textbf{undirected}.  If that is not the case, then the adjacency/weighting matrix is not symmetric and such graphs are called \textbf{directed graphs}.     
\bigskip

The above-introduced graph framework is quite general and admits application to many different scenarios. For example, when performing an opinion poll within a social network, the members of that social network are treated as the vertices (data acquisition points). Their friendship relations are represented by the edges which model graph connectivity while the member answers play the role of  graph signal values. 

{\color{black}The definition of an appropriate graph structure is a prerequisite for physically meaningful and computationally efficient
graph signal processing applications. Three important classes of problems, regarding the way how the graph topology is defined, are described in Sidebar 1.}

\textit{In the following, we shall demonstrate how this simple and intuitive  concept provides a natural and straightforward platform to introduce the graph-counterparts of several fundamental signal processing algorithms.}

\section{System on a Graph}

The signal shift operator (unit time delay) is the lynchpin in discrete-time signal processing, but it is not so obvious to define on graphs due to the rich underlying connectivity structure. Topologically, the signal shift on a graph can be viewed as the movement of a signal sample from the considered vertex along all edges connected to this vertex. The signal (backward) shift operator can then be compactly defined using the graph adjacency matrix as $\mathbf{x}_{shifted}=\mathbf{A}\mathbf{x}$.

\begin{figure}[t]
	\centering
	\includegraphics[trim = 80mm 100mm 0mm 0mm, clip,  width=0.46\columnwidth]{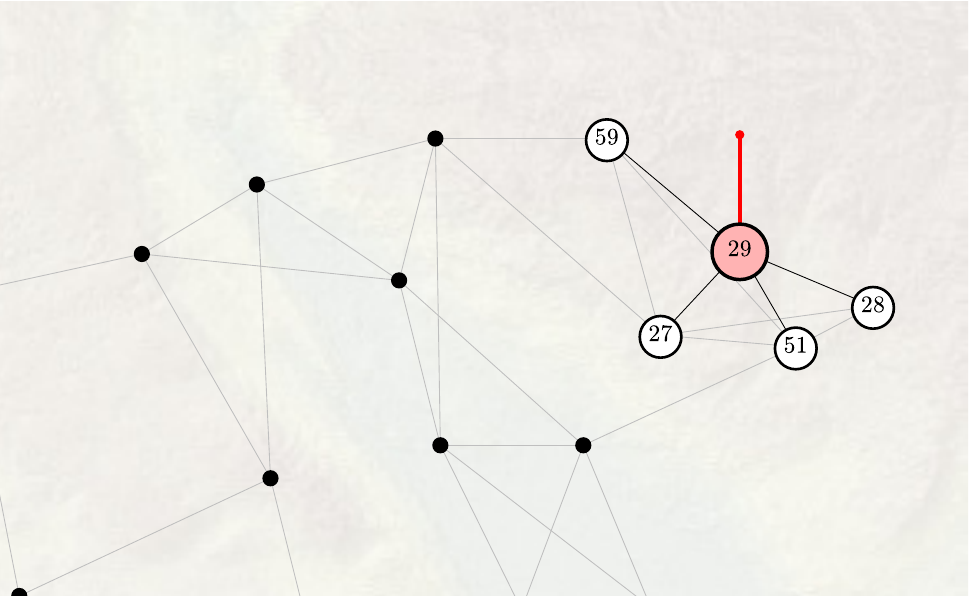}
		\hspace{2mm}
		\includegraphics[trim = 80mm 100mm 0mm 0mm, clip,  width=0.46\columnwidth]{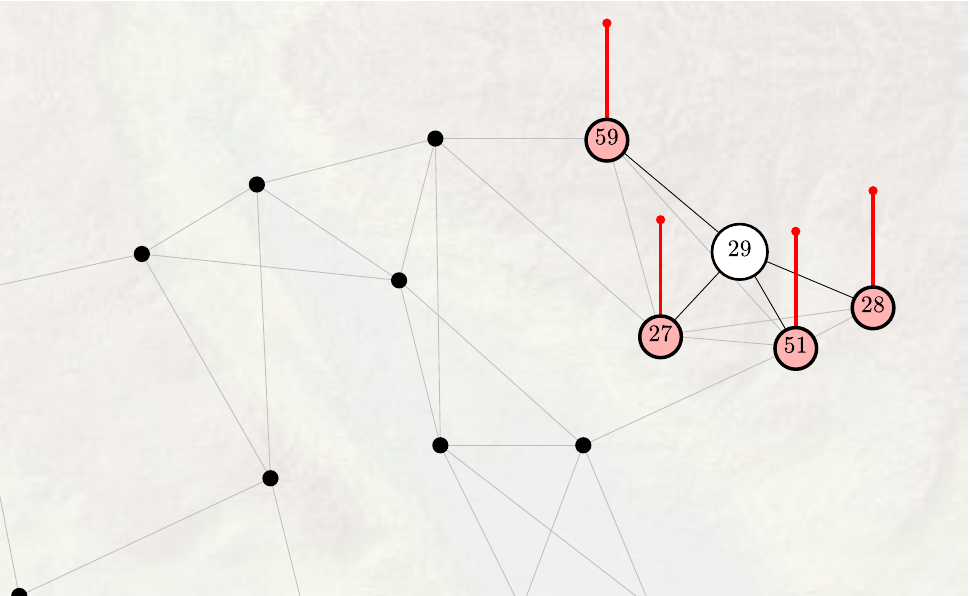}
		
	\caption{A single pulse graph signal $\mathbf{x}$ at the vertex $n=29$, that is, $x(n)=\delta(n-29)$ , and its graph shifted version $\mathbf{x}_{shifted}=\mathbf{A}\mathbf{x}$. The shift operator is demonstrated on the north-east part of the graph from Fig. \ref{fig:Prva_slika_e}, around the vertex $n=29$, is presented.}
	\label{fig:Prva_slika_shift}
\end{figure}

To draw distinction between the standard shift and the graph shift operator, consider the line graph in Fig. \ref{fig:Prva_slika_a}b) (bottom) and the ``spatial aware" graph in Fig. \ref{fig:Prva_slika_b}a), b), and assume that the input signal is a pulse that occurs only at the sensor $n=29$, that is, $x(n)=\delta (n-29)$. The shifted signal in classic signal processing (line graph in the bottom of Fig. \ref{fig:Prva_slika_a}b) (bottom)) will be  $x_{shifted}(n)=\delta(n-28)$ and can be considered as a movement of the delta pulse along the line graph from vertex $n$ to vertex $(n-1)$. The same principle can be applied to the graph domain in Fig. \ref{fig:Prva_slika_b}a) whereby the delta pulse from vertex $n=29$ is moved to all its connected vertices, to obtain the shifted graph signal,  $x_{shifted}(n)=\delta(n-27)+\delta(n-28)+\delta(n-51)+\delta(n-59)$, as shown in Fig. \ref{fig:Prva_slika_shift}. 

 If the shifted signal values are also scaled by the weighting coefficients of the corresponding edges, then the shifted signal is given by $\mathbf{W}\mathbf{x}$.  {\color{black} Since the Laplacian can also be used as a shift operator, we will adopt the symbol $\mathbf{S}$ to denote a general shift operator on a graph, which yields a graph shifted signal $\mathbf{S}\mathbf{x}.$}

\textit{Remark 3:} The standard shift operator, $x(n)=x(n-1)$, is a ``one-to-one" mapping, while the graph shift operator, $\mathbf{x}_{shifted}=\mathbf{S}\mathbf{x}$, is a ``one-to-many" mapping which accounts for the underlying physics of the sensing process {\color{black}(in our example)}, not possible to achieve with standard DSP. Moreover, it also allows us to incorporate a contextual relation between the vertices within the irregular grid trough the weighting matrix  $\mathbf{W}$.  \textit{Notice that the graph shift operator does not satisfy the isometry property since the energy of the shifted signal is not the same as the energy of the original signal. }

In analogy to the pivotal role of time shift in standard system theory, a system on a graph can be implemented as a linear combination of a graph signal and its graph shifted versions. {\color{black} The notion of a system is used in its classical sense, as a set of physical rules (an algorithm) that transforms an input graph signal into another (output) graph signal.} The output graph signal from a system on a graph can then be written as
\begin{equation}
\mathbf{y}=h_0 \mathbf{S}^0\, \mathbf{x}+h_1 \mathbf{S}^1\, \mathbf{x}+\dots+h_{M-1} \mathbf{S}^{M-1}\, \mathbf{x}=\sum_{m=0}^{M-1}h_m \mathbf{S}^m\, \mathbf{x},
\label{eq:filter-time}
\end{equation}
where,  by definition $\mathbf{S}^0=\mathbf{I}$, while $h_0$, $h_1$,  \ldots, $h_{M-1}$ are the system coefficients to be found (see Section \ref{SpecDFD}). Notice that for the directed and unweighted line graph in Fig. \ref{fig:Prva_slika_a}b) (bottom), the system on a graph in (\ref{eq:filter-time}) reduces to the well known standard Finite Impulse Response (FIR) filter, given by
\begin{equation}
y(n)=h_0x(n)+h_1x(n-1)+\cdots+h_{M-1}x(n-M+1).\label{ClassFIR}
\end{equation}

\textit{Remark 4:} The above established link between the classical transfer function of a physical system and its graph-theoretic counterpart may serve to promote new algorithmic approaches, which stem from signal processing, into many application scenarios that are directly considered as graphs. 

Observe that the Laplacian operator applied on a signal,  $\mathbf{L}\mathbf{x}$, can be considered as a combination of the scaled original signal, $\mathbf{D}\mathbf{x}$, and its weighted shifted version, $\mathbf{W}\mathbf{x}$, since  $\mathbf{L}\mathbf{x}=\mathbf{D}\mathbf{x}-\mathbf{W}\mathbf{x}$.
{\color{black}A system defined using the graph Laplacian is obtained from (\ref{eq:filter-time}) by replacing $\mathbf{S}=\mathbf{L}$, and has the form}
\begin{equation}
\mathbf{y}=\mathbf{L}^0\, \mathbf{x}+h_1 \mathbf{L}^1\, \mathbf{x}+\dots+h_{M-1} \mathbf{L}^{M-1}\, \mathbf{x}
\label{eq:filter-timeL}
\end{equation}
therefore allows us to always produce an unbiased estimate of a constant $c$, that is, if $\mathbf{x}=\mathbf{c}$ then $\mathbf{y}=\mathbf{c}$, since $\mathbf{L} \mathbf{c}=\mathbf{0}$.

From (\ref{eq:filter-timeL}), a simple first order system based on the graph Laplacian  can be written as
\begin{equation}
\mathbf{y}=\mathbf{x} + h_1 \mathbf{L} \mathbf{x}
\label{fist-order-system}
\end{equation}
and \textit{is amenable, with slight modifications, to being used for efficient low-pass graph filtering}, see Sidebar 2.  

\begin{figure*}[tbp]
\centering
\fboxsep=1pt
\fbox{\includegraphics[]{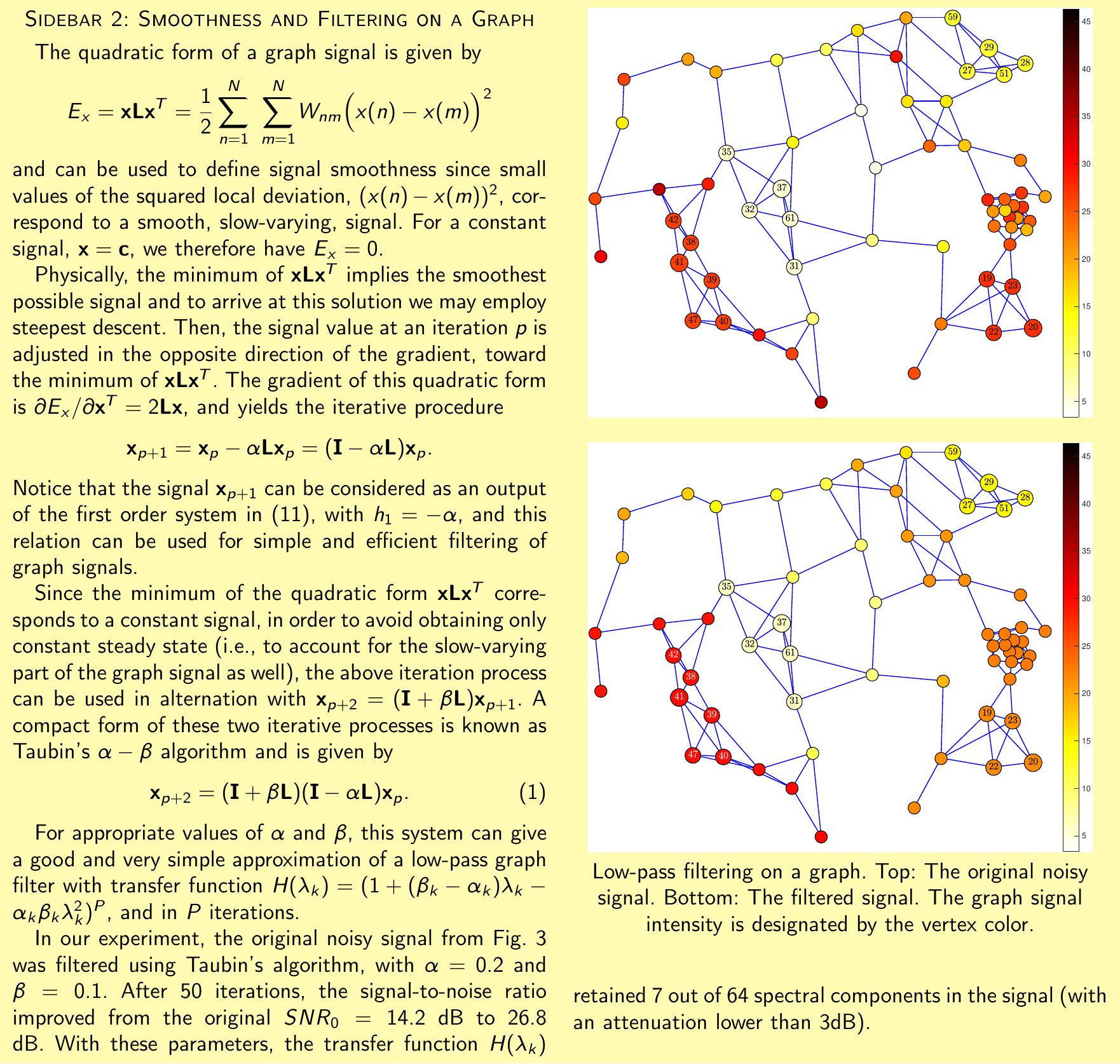}}
\end{figure*}

\bigskip
\textit{Remark 5:}  A system on a graph is conveniently defined by  the ``graph transfer function", {\color{black} $H(\mathbf{S})$, as   
\begin{equation}
\mathbf{y}=H(\mathbf{S})\mathbf{x}. \label{GSis}
\end{equation}

For an unweighted graph, the adjacency matrix, $\mathbf{A}$, is commonly used as a shift matrix, $\mathbf{S}$, while the Laplacian matrix, $\mathbf{L}=\mathbf{D}-\mathbf{W}$, is used to define a shift  on a weighted graph.}

\textbf{Properties of a system on a graph:} Following the above discussion, it is now possible to link the properties of linear systems with those of systems on a graph. From equations (\ref{eq:filter-time})-(\ref{GSis}) the system on a graph is said to be:
\begin{itemize}
\item Linear, if 
$$H(\mathbf{S})(a_1\mathbf{x}_1+a_2\mathbf{x}_2)=a_1\mathbf{y}_1+a_2\mathbf{y}_2.$$ 
\item Shift invariant, if
$$H(\mathbf{S})(\mathbf{S}\mathbf{x})=\mathbf{S}(H(\mathbf{S})\mathbf{x}).$$
\end{itemize}

\textit{Remark 6:} A system on a graph, defined by 
\begin{equation}
H(\mathbf{S})=h_0 \mathbf{S}^0+h_1 \mathbf{S}^1+\dots+h_{S-1} \mathbf{S}^{M-1} \label{Vetrx_DF}
\end{equation}
is linear and shift invariant, since the matrix multiplication of the square weighting matrices is associative $\Big( \mathbf{S}(\mathbf{S}\mathbf{S})=(\mathbf{S}\mathbf{S})\mathbf{S} \Big)$, that is  $\mathbf{S}\mathbf{S}^m=\mathbf{S}^m\mathbf{S}$.  
\bigskip

\section{Graph Fourier Transform}

While classic spectral analysis is performed in the Fourier domain,  spectral representations of graph signals employ either the adjacency/weighting matrix or the graph Laplacian eigenvalue decomposition. For the latter case we have
$$ \mathbf{L}=\mathbf{U}\mathbf{\Lambda} \mathbf{U}^{-1},$$
where $\mathbf{U}$ is an orthonormal matrix of the eigenvectors, $\mathbf{u}_k$, of the graph Laplacian matrix, $\mathbf{L}$, (in its columns), and $\mathbf{\Lambda}$ is a diagonal matrix of the corresponding eigenvalues, $\lambda_k$. These eigenvectors may then be used {\color{black}for the spectral-based clustering of graph vertices}, see Sidebar 3.

\begin{figure*}[tbp]
\centering
\fboxsep=1pt
\fbox{\includegraphics[]{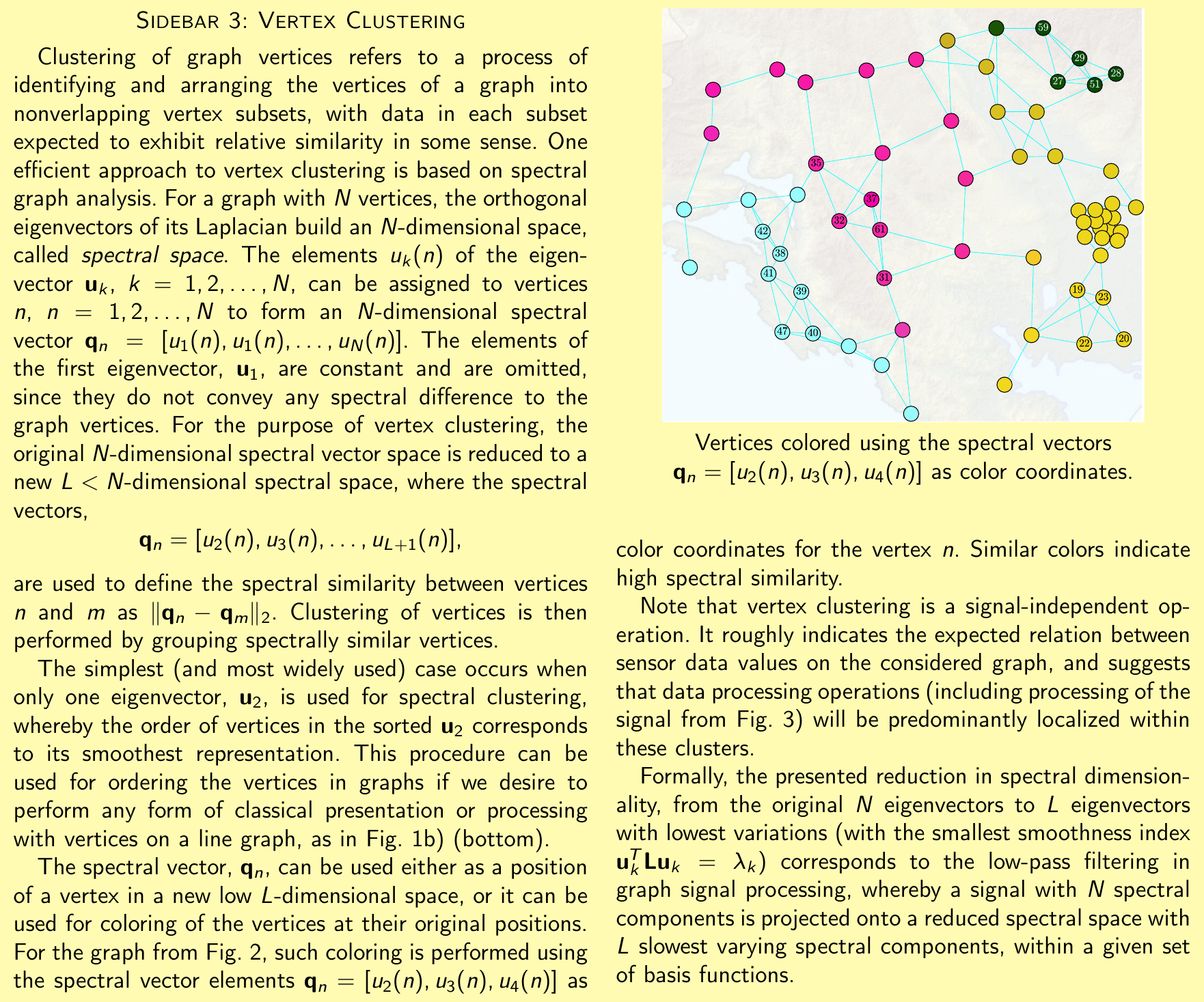}}
\end{figure*}

The graph Fourier transform, $\mathbf{X}$, of a graph signal, $\mathbf{x}$, is then defined as
\begin{equation}
\mathbf{X}=\mathbf{U}^{-1}\mathbf{x}.
\label{eq:gft}
\end{equation}

Physically, since $\mathbf{U}^{-1}=\mathbf{U}^{T}$,  the element $X(k)$ of {\color{black} a graph Fourier transform}, $\mathbf{X}$, represents a projection of the graph signal, $\mathbf{x}$, onto the $k$-th eigenvector, $\mathbf{u}_k\in \mathbf{U}$, that is
\begin{gather}
X(k)=\sum_{n=1}^{N}x(n)u_k(n). \label{GFTDEF}
\end{gather}

The {\color{black} inverse graph Fourier transform} is then straightforwardly obtained as
\begin{equation}
\mathbf{x}=\mathbf{U}\,\mathbf{X}
\label{eq:igft}
\end{equation}
or
\begin{gather}
x(n)=\sum_{k=1}^{N}X(k)u_k(n). \label{IGFTDEF}
\end{gather}

\textit{Remark 7:} In analogy to the  classic Fourier transform where the signal is projected onto a set of harmonic orthogonal bases, $\mathbf{X}=\mathbf{U}^{-1}\mathbf{x}$, where $\mathbf{U}$ is the matrix of harmonic bases $\mathbf{u}_k=[1, e^{j2\pi k/N}, \dots, e^{j\pi (N-1)k/N}]^T/\sqrt{N}$,  the graph  Fourier transform can be understood as a signal decomposition onto the set of eigenvectors of the graph Laplacian (or the adjacency matrix) that serve as orthonormal basis functions. In the case of a circular graph, the graph Fourier transform reduces to the standard discrete Fourier transform (DFT). For this reason, the transform in (\ref{GFTDEF}) is referred to as the Graph {\color{black} Fourier transform (GFT)}.

Classic spectral analysis can thus be considered as a special case of graph signal spectral analysis, with the adjacency matrix defined on an unweighted   circular directed graph (a line graph with the  connected last and first vertex), when $\mathbf{u}_k=[1, e^{j2\pi k/N}, \dots, e^{j\pi (N-1)k/N}]^T/\sqrt{N}$. This becomes obvious by recognizing that the eigenvalues of a directed unweighted circular graph,  $\lambda_k=e^{-j2\pi k /N}$, are easily obtained as a solution of the eigenvalue/eigenvector (EVD) relation $\mathbf{Au}_k=\lambda_k\mathbf{u}_k$. For a vertex $n$, this relation is of the form $u_k(n-1)=\lambda_k u_k(n)$. The previous vector elements $u_k(n)$ and eigenvalues $\lambda_k$ are the solutions of this difference equation. {\color{black} It can be shown that the eigenvectors of the graph Laplacian of a line graph are real-valued harmonic functions, whose combinations can produce the standard complex-valued DFT basis functions, in an indirect way.} The standard signal representation in Fig \ref{fig:Prva_slika_a}b) therefore corresponds to a signal whose domain is a line graph.

{\color{black}
As is common in signal processing, the true temperature was simulated through a linear combination of several graph Laplacian eigenvectors (serving as basis functions) in the form $\mathbf{x}=-160\mathbf{u}_1+16\mathbf{u}_2-8\mathbf{u}_3-40\mathbf{u}_4+ 16\mathbf{u}_5-24\mathbf{u}_6+\varepsilon(n),$ where the random Gaussian noise, $\varepsilon(n)$, had standard deviation  $\sigma_{\varepsilon}=4$.}

\section{Spectral Domain of a System on Graphs}
Consider a system on a graph, as in (\ref{eq:filter-timeL}), defined by its Laplacian matrix, given by
\begin{equation}
\mathbf{y}=\sum_{m=0}^{M-1}h_m \mathbf{L}^m\, \mathbf{x}.
\label{eq:filter-lap}
\end{equation}
Upon employing the eigen-domain (graph spectral) representation of the Laplacian matrix, $\mathbf{L}=\mathbf{U}\mathbf{\Lambda}\mathbf{U}^{-1}$, we have
\begin{equation}
\mathbf{y}=\sum_{m=0}^{M-1}h_m \mathbf{U} \mathbf{\Lambda}^m \mathbf{U}^{-1}\, \mathbf{x}
=
\mathbf{U}\, H(\mathbf{\Lambda})\mathbf{U}^{-1}\, \mathbf{x},
\label{eq10}
\end{equation}
where
\begin{equation}
H(\mathbf{\Lambda})=
\sum_{m=0}^{M-1}h_m \mathbf{\Lambda}^m
\label{tf}
\end{equation}
is the transfer function of the graph system.

From (\ref{eq10}),
$
\mathbf{U}^{-1}\mathbf{y}=
 H(\mathbf{\Lambda})\mathbf{U}^{-1}\, \mathbf{x}
$,
or in terms of the graph Fourier transform of the input and output signal
 \begin{equation}
\mathbf{Y}=
 H(\mathbf{\Lambda})\, \mathbf{X}.
\end{equation}

The classic spectral transfer function for (\ref{ClassFIR}) is then obtained by using the adjacency matrix of an unweighed directed circular graph whose eigenvalues are  $\lambda_k=e^{-j2\pi k /N}$.

\section{Spectral Domain Filter Design}\label{SpecDFD}

Consider a desired  graph transfer function, $G(\mathbf{\Lambda})$. Like in classic signal processing, a system with this transfer function can be implemented either in the spectral domain or in the vertex domain. 

The spectral domain implementation is straightforward and can be performed in the following three steps: 
\begin{enumerate}
\item 
Calculate the {\color{black} GFT} of the input graph signal $\mathbf{X}=\mathbf{U}^{-1}\mathbf{x}$, 
\item 
Multiply the {\color{black} GFT} of the input graph signal with transfer function  $G(\mathbf{\Lambda})$ to obtain  $\mathbf{Y}=G(\mathbf{\Lambda})\mathbf{X}$, and 
\item 
Calculate the output graph signal as the
inverse graph Fourier transform of $\mathbf{Y}$ to yield  $\mathbf{y}=\mathbf{U}\mathbf{Y}$. 
\end{enumerate}

Notice that this procedure may be computationally very demanding for large graphs where it may be easier to implement the desired filter (or its close approximation) in the vertex domain, in analogy to the time domain in the classical approach. This means that we have to find the coefficients, $h_0,h_1,\dots,h_{M-1}$ in (\ref{eq:filter-time}), such that its spectral representation, $H(\mathbf{\Lambda})$,  is equal (or at least as close as possible) to the desired $G(\mathbf{\Lambda})$.

\begin{figure*}[tbp]
\centering
\fboxsep=1pt
\fbox{\includegraphics[]{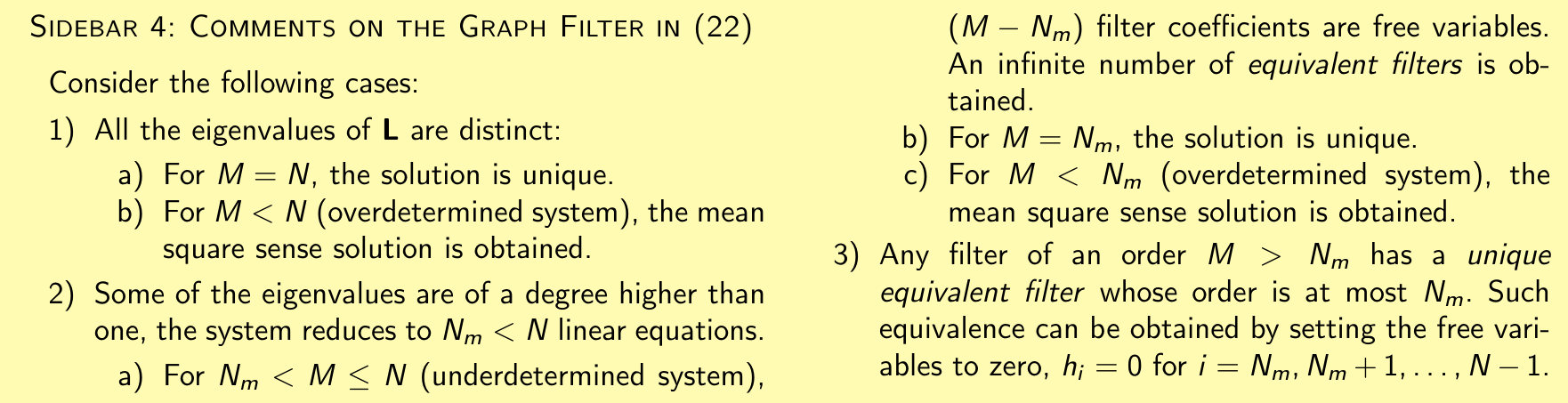}}
\end{figure*}

In other words, the transfer function of the vertex domain system in (\ref{tf}), given by
$H(\lambda_k)=h_0 +h_1 \lambda_k^1+\dots h_{M-1} \lambda_k^{M-1}$, should be equal to the desired transfer function, $G(\lambda_k)$, for each spectral index, $k$. This condition leads to a system of linear equations
\begin{gather}
	h_0 +h_1 \lambda_1^1+\dots h_{M-1} \lambda_1^{M-1}=G(\lambda_1) \nonumber \\
	h_0 +h_1 \lambda_2^1+\dots h_{M-1} \lambda_2^{M-1}=G(\lambda_2) \nonumber
	\nonumber \\
	\vdots \nonumber \\
	h_0 +h_1 \lambda_{N}^1+\dots h_{M-1} \lambda_{N}^{M-1}=G(\lambda_{N}).
	\label{eq:sistem-filt}
\end{gather}
The matrix form of this system is given by 
\begin{equation} 
\mathbf{V}\!_\lambda \, \mathbf{h}=\mathbf{g}, \label{eq:sistem-filtV}
\end{equation}
where $\mathbf{V}\!_\lambda$ is a Vandermonde matrix formed of the eigenvalues, $\lambda_k$, while
$\mathbf{h}=[h_0, h_1,\dots,h_{M-1}]^T$
is the vector of system coefficients that we wish to estimate, and   
$$\mathbf{g}=[G(\lambda_1), G(\lambda_2),\dots,G(\lambda_{N})]^T=\textrm{diag}(G(\mathbf{\Lambda})).$$

The system order $M$ is typically significantly lower than the number of equations, $N$, in (\ref{eq:sistem-filt}).
For such an overdetermined case, the least-squares approximation  of $\mathbf{h}$ is obtained by minimizing the squared error, 
$ e^2=\left\Vert \mathbf{V}\!_\lambda \mathbf{h}-\mathbf{g}\right\Vert _2 ^2.$
Like in standard least-squares, the solution is obtained by a direct minimization, $\partial e ^2 /\partial \mathbf{h}^T=\mathbf{0} $, to yield 
\begin{equation}
\mathbf{\hat h}=(\mathbf{V}_{\!\lambda}^T \mathbf{V}_{\!\lambda})^{-1}\mathbf{V}_{\!\lambda}^T \mathbf{g}=\text{pinv}(\mathbf{V}\!_\lambda)\mathbf{g}. \label{MSAPP}
\end{equation} 
The so obtained solution, $\mathbf{\hat h}$, therefore represents the mean square error minimizer for $ \mathbf{V}\!_\lambda \mathbf{h}=\mathbf{g}$. Notice that this solution may not satisfy $ \mathbf{V}\!_\lambda \mathbf{h}=\mathbf{g}$, in which case the coefficients $\mathbf{\hat g}$ (its spectrum $\hat{G}(\mathbf{\Lambda})$) may be used, that is
$$ \mathbf{V}\!_\lambda \mathbf{\hat{h}}=\mathbf{\hat{g}}.$$
Such a  solution, in general, differs from the desired system coefficients $\mathbf{g}$ (its spectrum $G(\mathbf{\Lambda})$).

\textit{Example:} Consider the graph signal from Fig.~\ref{fig:Prva_slika_e}. The task is to design a graph filter whose frequency response is $g(\lambda_k)=\exp(-\lambda_k)$ and to then filter the graph signal using this spectral domain graph filter. For $M=4$,  the corresponding system coefficients can be found to be
$h_0=0.9606$,
$h_1=-0.7453$,
$h_2=0.1936$, and
$h_3=-0.0162$. Upon signal filtering using the so defined graph transfer function,  the output signal-to-noise ratio was $SNR=21.74$  dB, that is a $7.54$ dB improvement over the original  signal-to-noise ratio  $SNR_0=14.2$  dB.    

More detail on the solution of the system in (\ref{eq:sistem-filt}) and (\ref{eq:sistem-filtV}) is provided in Sidebar 4.

\section{Optimal Denoising}\label{OptimaDenoising}

Consider a measurement, as in the temperature measurement scenario in Fig. \ref{fig:Prva_slika_a}, which is composed of a slow-varying desired signal, $\mathbf{s}$, and a superimposed fast changing disturbance, $\boldsymbol{\varepsilon}$, to give
$$
\mathbf{x}=\mathbf{s}+\boldsymbol{\varepsilon}.
$$
The aim is to design a graph filter for disturbance suppression (denoising), the output of which is denoted by $\mathbf{y}$, \cite{Opt}.

The optimal denoising task can then be defined through a minimization of the cost function
\begin{equation}
J=\frac{1}{2}\Vert \mathbf{y}-\mathbf{x}\Vert_2^2 + \alpha \mathbf{y}^T \mathbf{L} \mathbf{y}.
\label{minim}
\end{equation}
The minimization of the first term, $\frac{1}{2}\Vert \mathbf{y}-\mathbf{x}\Vert_2^2$, enforces the output signal, $\mathbf{y}$, to be as close as possible, in terms of the minimum residual disturbance power, to the available observations, $\mathbf{x}$. As mentioned before, the second term, $\mathbf{y}^T \mathbf{L} \mathbf{y}$, represents a measure of smoothness of the graph filter output,  $\mathbf{y}$. For more detail on promoting smoothness of a graph signal, see Sidebar 2. The parameter $\alpha$ models a balance between the closeness of the output, $\mathbf{y}$, to the observed data, $\mathbf{x}$,  and the smoothness of output estimate $\mathbf{y}$. While the problem in (\ref{minim}) could be expressed through a constrained Lagrangian  optimization, whereby we choose to focus more on the graph theoretic issues and hence we adopt a simpler option whereby the mixing parameter $\alpha$ is chosen empirically.

The solution to this minimization problem follows from
$$
\frac{\partial J}{\partial \mathbf{y}^T} = \mathbf{y}-\mathbf{x} +  2\alpha \mathbf{L} \mathbf{y} =\mathbf{0}
$$
and results in a smoothing optimal denoiser in the form
$$
\mathbf{y}= (\mathbf{I} + 2\alpha \mathbf{L})^{-1} \mathbf{x}.
$$
The Laplacian spectral domain form of this relation is
$$
\mathbf{Y}= (\mathbf{I} + 2\alpha \mathbf{\Lambda})^{-1} \mathbf{X}, 
$$
with the corresponding graph filter transfer function
$$
H(\lambda_k)=\frac{1}{1+2\alpha\lambda_k}.
$$
For a small $\alpha$, $H(\lambda_k) \approx 1$ and $\mathbf{y} \approx \mathbf{x}$, while for a large $\alpha$, $H(\lambda_k) \approx \delta(k)$ and $ \mathbf{y} \approx const.$, which enforces $\mathbf{y}$ to be maximally smooth (a constant, without any variation). Using $\alpha=4$, the obtained output signal-to-noise ratio for the graph signal from  Fig.~\ref{fig:Prva_slika_e}  was $SNR=26$ dB, a $11.8$ dB improvement over the original $SNR_0=14.2$ dB.

\textit{Remark 8:} There are many cases when the graph topology is unknown, so that the graph structure, i.e.,
the Laplacian (graph edges and their weights) is also unknown. 
To this end, we may employ a class of methods for graph topology learning, based on the minimization of the cost function in (\ref{minim}) with respect to both the Laplacian, $\mathbf{L}$, and the output signal, $\mathbf{y}$, with additional (commonly sparsity) constraints imposed on the Laplacian values.

\begin{figure*}[tbp]
	\centering
	\fboxsep=1pt
	\fbox{\includegraphics[]{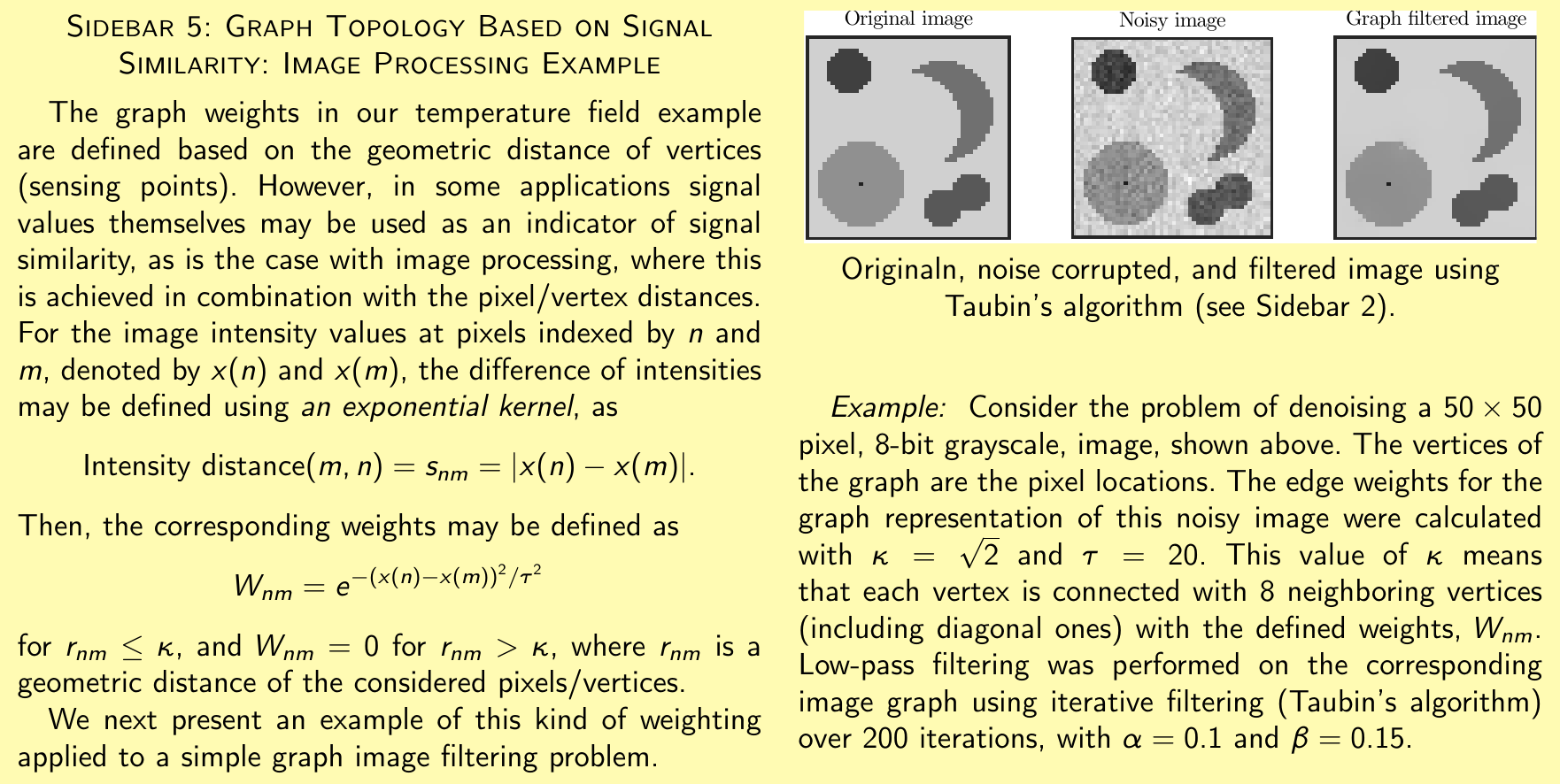}}
\end{figure*}

\section{Current Graph Signal Processing Challenges}

{\color{black}
Current research is mainly focused on graphs themselves, like for example,  on reducing the complexity of calculation in very large graphs, including downsampling, multirate analysis, compressive sensing, graph segmentation,  non-linear GSP, robust GSP, deep learning architectures for graph signals, multidimensional graph signals, and vertex-varying and vertex-frequency analysis. 
}

\section{What We Have Learned}

Natural signals (speech, biomedical, video) reside over irregular domains and are, unlike the signals in communications, not adequately processed using, e.g., standard harmonic analyses. 
While Data Analytics are heavily dependent on advances in DSP, neither the EE graduates worldwide nor practical data analysts are yet best prepared to employ graph algorithms in their future jobs. Our aim has been to fill this void by providing an example-driven platform to introduce graphs and their properties through the well understood notions of transfer functions, Fourier transform, and digital filtering. 

While both a graph with $N$ vertices and a classical discrete time signal with $N$ samples can be viewed as $N$-dimensional vectors, structured graphs are much richer irregular domains which convey information about both the signal generation and propagation mechanisms. This allows us to employ intuition and our know-how from Euclidean domains to revisit basic dimensionality reduction operations, such as coarse graining of graphs (\textit{cf.} standard downsampling). In addition, in the vertex domain a number of different distances (shortest-path, resistance, diffusion) have useful properties which can be employed to maintain data integrity throughout the processing, storage, communication and analysis stages, as the connectivities and edge weights are either dictated by the physics of the problem at hand or are inferred from the data. This particularly facilitates maintaining control and intuition  over distributed operations throughout the processing chain.

It is our hope that this lecture note has helped to
demystify graph signal processing for students and educators, together with empowering practitioners with enhanced intuition in graph-theoretic  design and optimization. This material may also serve as a vehicle to seamlessly merge curricula in Electrical Engineering and Computing. 
The generic and physically meaningful nature of this example-driven Lecture Note is also likely to promote intellectual curiosity and serve as a platform to explore the numerous opportunities in manifold applications in our ever-growing interconnected world, facilitated by the Internet of Things.

\section*{Acknowledgments} We are privileged to have had the help and advice of one of the pioneers in Graph Theory Professor Nicos Christofides. We are grateful for his time, his incisive comments and valuable advice. We would also like to express our  sincere gratitude to the students in our respective postgraduate courses, for their feedback on the material taught based on this Lecture Note.

\section*{Authors}

\textbf{Ljubi\v{s}a Stankovi\'{c}}, FIEEE, (ljubisa@ac.me) is professor at the 
University of Montenegro. His research interests include time-frequency analysis, compressive sensing, and graph signal processing.  He is a vice-president of the National Academy of Sciences and Arts of Montenegro (CANU) and a member of the European Academy of Sciences and Arts. Prof. Stankovi\'{c} is a recipient of the 2017 EURASIP Best Journal Paper Award.

\textbf{Danilo P. Mandic}, FIEEE, (d.mandic@imperial.ac.uk) 
is a professor of signal processing at Imperial College London, United Kingdom. He is a member of the IEEE Signal Processing Society Education Technical Committee, and has received President’s Award for Excellence in Postgraduate Supervision at Imperial College. He is a recipient of the 2018 Best Paper Award in IEEE Signal Processing Magazine.

\textbf{Milo\v{s} Dakovi\'{c}} (milos@ac.me) is professor at
the University of Montenegro. His research interests include
graph signal processing, and time-frequency analysis. 

\textbf{Ilya Kisil} (i.kisil15@imperial.ac.uk) is a Ph.D. candidate at Imperial College London. His research interests include tensor decompositions, big data, efficient software for large scale problems, and graph signal processing. 

\textbf{Ervin Sejdi\'{c}}, SMIEEE, (esejdic@ieee.org) is an assistant
professor at the University of Pittsburgh, USA. His research interests
include biomedical signal processing, rehabilitation
engineering, and neuroscience. He received the USA
Presidential Early Career Award for Scientists and Engineers in
2016.

\textbf{Anthony G. Constantinides}, LFIEEE, (a.constantinides @imperial.\-ac.uk) 
is emeritus professor of signal processing in the Department of Electrical and Electronic Engineering at Imperial College London, United Kingdom.

\end{document}